\documentclass{article}
\usepackage{graphics}
\usepackage[dvips]{graphicx}
\usepackage{bm}
\usepackage{latexsym}
\usepackage{pxfonts}
\begin{document}

\title{Tunable variation of optical properties of polymer capped gold  nanoparticles}

\author{M. Haridas, S. Srivastava and J. K. Basu\\
\emph{\small{Department of Physics, Indian Institute of Science, Bangalore, 560012, India}}}

\maketitle
\abstract{
Optical properties of polymer capped gold nanoparticles of various sizes (diameter 3-6 nm) have been studied. We present a new scheme to extract size dependent variation of total dielectric function of gold nanoparticles from measured UV-Vis absorption data. The new scheme can also be used, in principle, for other related systems as well. We show how quantum effect, surface atomic co - ordination and polymer - nanoparticle interface morphology leads to a systematic variation in inter band part of the dielectric function of gold nanoparticles, obtained from the analysis using our new scheme. Careful analysis enables identification of  the possible changes to the electronic band structure in such  nanoparticles. 

\section{Introduction}
\label{intro}
Optical properties of metal nanoparticles, especially noble metals, have been experimentally investigated over the last couple of decades[1-15], Initial studies\cite{8,16,17} were mostly performed on nanoparticles prepared by thermal evaporation, and were without any protective coating. The basic interest was from the point of view of understanding finite size effects on the optical properties and their possible interpretation in terms of the modification of electronic band structure. The rapid rise of the field of nano material science has lead to various controlled methods of growth of such nanoparticles\cite{18}, especially gold and silver, which have also been passivated for application in various fields like sensors  etc. It turns out that the process of capping could lead to strong modifications of various physical properties like optical[19-22], magnetic\cite{23}, electronic \cite{24,25} and thermal\cite{26}.
      Most of the studies which have been performed on optical properties of metal nanoparticles have concentrated on the finite size and, more recently, surface effects on the nature of the plasmon resonance. It was shown\cite{7}, how the standard Mie\cite{1,2} absorption coefficient using a dielectric function based on the simple Drude model of free electrons is unable to explain the position as well as width of the resonances in both gold and silver nanoparticles. Inclusion of contribution of inter band absorption in to the dielectric function of the  nanoparticles, especially for gold, bring the modified - Mie calculation and data closer, but still the discrepancies are usually quite large. Except for few isolated studies\cite{13}, not much attention has been paid to modeling the observed absorption spectra for gold in the UV region, away from the plasmon resonance. In any case, it has always been assumed that finite size and surface effects leads to modification of optical properties, which can be modeled by introducing a size dependent damping term in the free electron    dielectric function, leaving the inter band contribution to dielectric function unchanged. Early work \cite{16,17} on metal cluster films did provide some evidence for size dependent modification of electronic band structure (mostly d bands in noble metals), using Ultra violet - Visible (UV-Vis) absorption and Xray Photo electron Spectroscopy (XPS) measurements. However there is some controversy \cite{8} about the interpretation of the  optical properties of such metal cluster films as it was not clear to what extent the observed spectra depended on the morphology of the clusters. In addition, no clear trends exists in terms of width and  position of the centroid of the valence bands as a function of size and surface effects\cite{27,28}. Some effort was made to calculate  the optical properties of metal clusters by modeling the inter band part of dielectric function with a single Lorentzian oscillator. However the predictions made were qualitative  in nature and no effort was made to match the experimentally obtained dielectric functions of metals. However a recent work \cite{29} has provided an analytical model for the most reliable experimental data on dielectric function of bulk gold\cite{9}. This allows modeling of total dielectric function of gold nanoparticles.
 
                         \hspace{0.4in}
 We present results on UV-Vis optical spectroscopy  of polymer - capped gold nanoparticles of various sizes (3 - 6 nm) by systematic variation of density of capping agent in the nanoparticles. We have used standard Mie theory, with modifications suggested by Kreibig et al\cite{30}, to calculate the optical spectra for gold nano particles, in solution, capped with  polymethyl methacrylate (PMMA). We present a new scheme for detailed analysis of the optical absorption data from gold nanoparticles whereby size dependent variation of dielectric function corresponding to both intraband (free electron) and interband (5 d to 6 s p) term can be extracted.  In the  above calculations,  we have used the analytical model\cite{29} for the dielectric function of gold and by fitting the obtained absorption spectra, extracted the size dependent dielectric function of the nanoparticles,  including the interband part, by varying the relevant parameters in the model. Our results indicate systematic variation in the  imaginary part of the dielectric function of gold nano particles as a function of capping concentration and size. We find that the peaks in the obtained imaginary dielectric function, corresponding to inter band transitions are either blue/red shifted or unchanged, with respect to the bulk values, as a function of capping concentration and size. The extent of the observed shifts was found to depend sensitively on the nature of the capping agent. Interestingly, powder X-ray diffraction (XRD) measurements on nanoparticles indicates slight lattice contraction, as compared to bulk gold, for the largest particles, studied here, while no significant change was observed for smaller sized particles and high capping concentrations. We have provided possible explanations for these observed behavior in terms of modifications to the band structure of polymer capped gold nanoparticles, arising out of  lattice contraction, reduced surface atomic co-ordination and polymer - nanoparticle interface morphology.

                  The remainder of our paper is arranged as follows. In section II we describe  our experimental methods, section III describes optical property of metal particles and gold nanoparticles in section IV followed by result and discussion in section V and VI, respectively. 
\section{Experimental Details}
\label{expdetal}

The PMMA capped gold nano particles used for this experiment were synthesized by reduction of HAuCl$_4$ solution in ethanol with freshly prepared aqueous solution of NaBH$_4$ in presence of PMMA, as described earlier\cite{26,40}. To control the size of the gold nanoparticles as well as vary the density of PMMA chains on the surface of the nanoparticles, the concentration of PMMA ($\mu$) was varied by a factor of 16.   The size of the nanoparticle formed is inversely proportional to PMMA capping concentration $\mu$\cite{26}.  The powders were extracted from solution by controlled evaporation for XRD (PANalytical ) measurements which were carried out with Cu K$_\alpha$ 1.54 $\AA$ X-rays from 35 $^o$ to 90 $^o$. The size and polydispersity were quantitatively estimated from transmission electron microscopy (TEM, Tecnai F 30) images of nano particles, obtained by application of a droplet of nano particle solution on carbon coated copper grids. The UV -Vis absorption measurements were carried out (GBC Cintra 40) in the range 250 nm to 650 nm. 
\section{Optical Absorption Spectroscopy of Small Metal Particles}
\label{optical absorption}
Noble metal nanoparticles exhibits pronounced resonance absorption of visible light due to collective excitation of quasi-free electrons often called  surface plasmon resonance. Mie was the first to obtain the exact solution of light absorption by a sphere in the frame of classical electrodynamics. The exact analytical solution exists only for some simple shapes of particles like sphere, ellipsoid. The main assumptions of Mie's theory of optical absorption by small particles is that the particle and its surrounding medium are each homogeneous and describable by their bulk optical dielectric functions.  In general, solutions to Maxwell's equations for this geometry yield an expression\cite{1,2}  for absorption has the form  $\sigma$=4$\pi$/k$^2$$\Sigma$(a$_n$$^2$+b$_n$$^2$),
where K is wave vector in the surrounding medium, a$_n$ and b$_n$ represents the electric and magnetic multipoles, which is expressed in terms of legendre and Bessel polynomials.  
 When the size of a particle is much smaller than the wavelength of the exciting radiation, the absorption is dominated by dipole term and  Mie's formula simplifies to the following form,
 \begin{equation}
{\cal \sigma } = 9\epsilon_m^{3/2}v\frac{\rm\omega}{\rm c}\frac{\rm\epsilon_2}{\rm (\epsilon_1+2\epsilon_m)^2+(\epsilon_2)^2}.
\end{equation}

Here $\epsilon_2$ and $\epsilon_1$ are imaginary and real part of dielectric constant $\epsilon$ (=$\epsilon_1$+i$\epsilon_2$) of the particle, $v$ is the volume,  $\epsilon_m$, is the dielectric constant of medium, c is the velocity  and $\omega$ is the frequency of light. Drude's free electron theory gives the dielectric constant in terms of the bulk plasma frequency $\omega_p$, as $\epsilon_f$=1-$\frac{\rm\omega_p^2}{\rm\omega^2+i\omega\gamma}$,
 where $\gamma$ is the damping constant. However it was found earlier that the width and position of the plasmon resonance in nanoparticles does depend on size \cite{30,32,33,34}. As size decreases the width of the plasmon resonance increases, while the absorption peak  shifts towards lower wavelength. Mie theory is obviously not able to explain these observed phenomena. In particles smaller than the mean free path of conduction electrons in bulk metal, the relaxation can be dominated by collision at the surface. The mean free path limitation has been discussed by Kreibig\cite{7}, who considered that the damping constant is increased due to additional collisions with the boundary of the particle, and it can be written as,\vspace{0.1in}
\begin{equation}
 \gamma(r) = \gamma_{bulk}+Av_f/r ,  
\end{equation}

where v$_f$ is the electron velocity at the Fermi surface, and $A$ is an emperical constant that includes details of the scattering process.  
\section{Optical Properties of Bulk and Nanoparticulate Gold}
\label{optbulknano}
The experimental data of Johnson and Christy\cite{9} on optical properties of gold, has been widely used as a reference for the dielectric function of gold in the range of 1-6 eV. There are two dominant contributions to the dielectric function in this region - the Drude free electron term and the dielectric function corresponding to interband absorption. Using the standard free electron dielectric function, the interband dielectric absorption contribution to the dielectric function can be evaluated using the total dielectric function as obtained by Johnson and Christy. However it is well known that both for gold and copper and to a lesser extent for silver, the absorption calculated based on Mie theory using size dependent damping term, under estimates the width of the plasmon resonance. This indicates presence of additional damping mechanism in such particles. It has also been observed that the absorption for gold nano particles, increases strongly in the high energy part of the spectrum above the plasmon resonance. This is due to the presence of strong interband (d - s) absorption in the visible and UV part of the spectrum. It  has been shown how by introducing an additional term in the nanoparticle dielectric function, corresponding to interband absorption, calculated absorption especially the plasmon resonance width is much closer to actual data. 
The modified\cite{30} Mie theory considering the complex interband  dielectric function of particle can be written  as, \vspace{0.1in}

\begin{equation}
{\cal \sigma} = \frac{\rm 9v}{\rm c}\frac{\rm\epsilon_m^{3/2}}{\rm 1+\epsilon_I^{R}+2\epsilon_m} \frac{\rm\Omega^2\gamma\omega^2+k\omega^3(\omega^2+\gamma^2)}{\rm (\Omega^2-\omega^2+k\omega\gamma)^2+(\gamma\omega+k\omega^2)^2},
\end{equation}
where
\begin{equation}
  \Omega = \frac{\rm\omega_p}{\rm (1+\epsilon_I^{R}+2\epsilon_m)^{1/2}}, 
\end{equation}
and
 \begin{equation}
      k=\frac{\rm\epsilon_I^{Im} }{\rm1+\epsilon_I^{R}+2\epsilon_m }.
\end{equation}

Here $\epsilon_I^R$ and $\epsilon_I^{Im}$ is the real and imaginary contributions to interband dielectric constants. The other parameters have the usual meaning.\vspace{0.1in}
It might be noted that, although the inter-band contribution $\epsilon_I$ is assumed to be independent of size,
the use of the above equation leads to satisfactory fit to optical absorption data on gold and silver nanoparticles in the region around the plasmon peak. However, not much attention is normally paid to the region in the absorption spectrum below the plasmon resonance wavelength which is dominated by the interband absorption from the d-bands to the sp like conduction bands. These effects are very strong for copper and gold since the absorption onset is $\sim$ 2 eV for both, which also happens to overlap with the their respective plasmon resonance, while it is not so dominant for silver for which the interband absorption onset is $\sim$ 4 eV. However, the interband absorption effect is so strong in copper that it tends to mask the plasmon resonance in optical measurements\cite{12}. Hence, gold seems to be most ideal system to study the effect of quantum confinement and surface effects on interband absorption. Except for some work by Yamaguchi et al \cite{17}, and some more recent work on lead and tin clusters\cite{27,35}, systematic study has not been performed on the size dependance of interband dielectric function, mainly due to lack of analytical expression ( equivalent to Drude free electron term ) describing it's behavior. Recently, an analytical  model for the total dielectric function ($\epsilon$), has been proposed and can be re-written as, 
\begin{equation}
 {\cal \epsilon } = \epsilon_\infty-\frac{\rm\omega_p^2}{\rm(\omega^2+i\gamma\omega)}+\sum_{i=1}^2 A_i\lbrack\frac{\rm \omega_i e^{i\Phi_i}}{\rm\omega_i-\omega-i\gamma_i}+\frac{\rm \omega_ie^{-i\Phi_i}}{\rm\omega_i+\omega+i\gamma_i}\rbrack ,    
 \label{1} 
\end{equation}

 where $\omega$ is the frequency of light, $\omega_1$ and $\omega_2$represents the frequencies corresponding to two interband transition. A$_1$ and A$_2$ dimensionless critical point amplitudes, $\gamma_1$and $\gamma_2$ the transition broadening expressed as frequency, $\omega_p$ is plasma frequency. The phases $\Phi_1$ and $\Phi_2$  have the value $\sim$ $\pi$/4. This matches Johnson and Christy data fairly closely.

Since the dielectric function is defined in terms of a set of physically meaningful parameters, it should be possible to obtain variation of dielectric function of gold nanoparticles, with possible extension to other similar material, due to various surface effects, which can sometimes be so strong  as to overwhelm the intrinsic quantum confinement effects. Since the dielectric function, specifically the imaginary part, can be related to the electronic band structure, it allows formulation of a very powerful method of extracting band structure modification in nanoparticles in general and gold nanoparticles in particular, due to quantum confinement and surface effects. It is well known that for gold there are two major interband transitions which contribute to the dielectric function below 5 eV. Hence we can rewrire the interband contribution to the dielectric function of gold as defined in Eq. 6 as  
\vspace{0.1in}

 \begin{equation}
 {\cal \epsilon }= \epsilon_\infty-\frac{\rm\omega_p^2}{\rm(\omega^2+i\gamma\omega)}+\sum_{i=1}^2 \lbrack e_i^R+ie_i^{Im}\rbrack .
\label{2}
\end{equation}
Here e$_i^R$ and e$_i^{Im}$ is the real and imaginary part, respectively, of the two interband terms contributing to the dielectric function, where $\epsilon_I^R$=$\sum$$e_i^R$ and $\epsilon_I^{Im}=\sum e_i^{Im}$. e$_i^{Im}$ can be written as  

\begin{equation}
 \small{e_i^{Im} =
      \frac{\rm A_i \omega \lbrack 2\omega_i cos(\phi_i) (\omega_i^2-\omega^2+\gamma_i^2)+4\omega \gamma_i (\omega_i  cos(\phi_i) - \gamma_i sin(\phi_i)) \rbrack}{\rm (\omega_i^2-\omega^2+\gamma_i^2)^2+ 4\gamma_i^2\omega^2}
 .} \label{eq:wideeq}
\end{equation}

In the analytical model (Eq.6) used here to represent the interband term, the two transitions are identified with energies $\omega_1$ ( = 2.65 eV) and  $\omega_2$ ( = 3.75 eV). From Eq.8, this corresponds to two resonant terms with peaks at  e$_1^{Im}$= 2.89 eV and  e$_2^{Im}$= 4.25 eV, in bulk gold.
\begin{figure}
\includegraphics[scale=0.6]{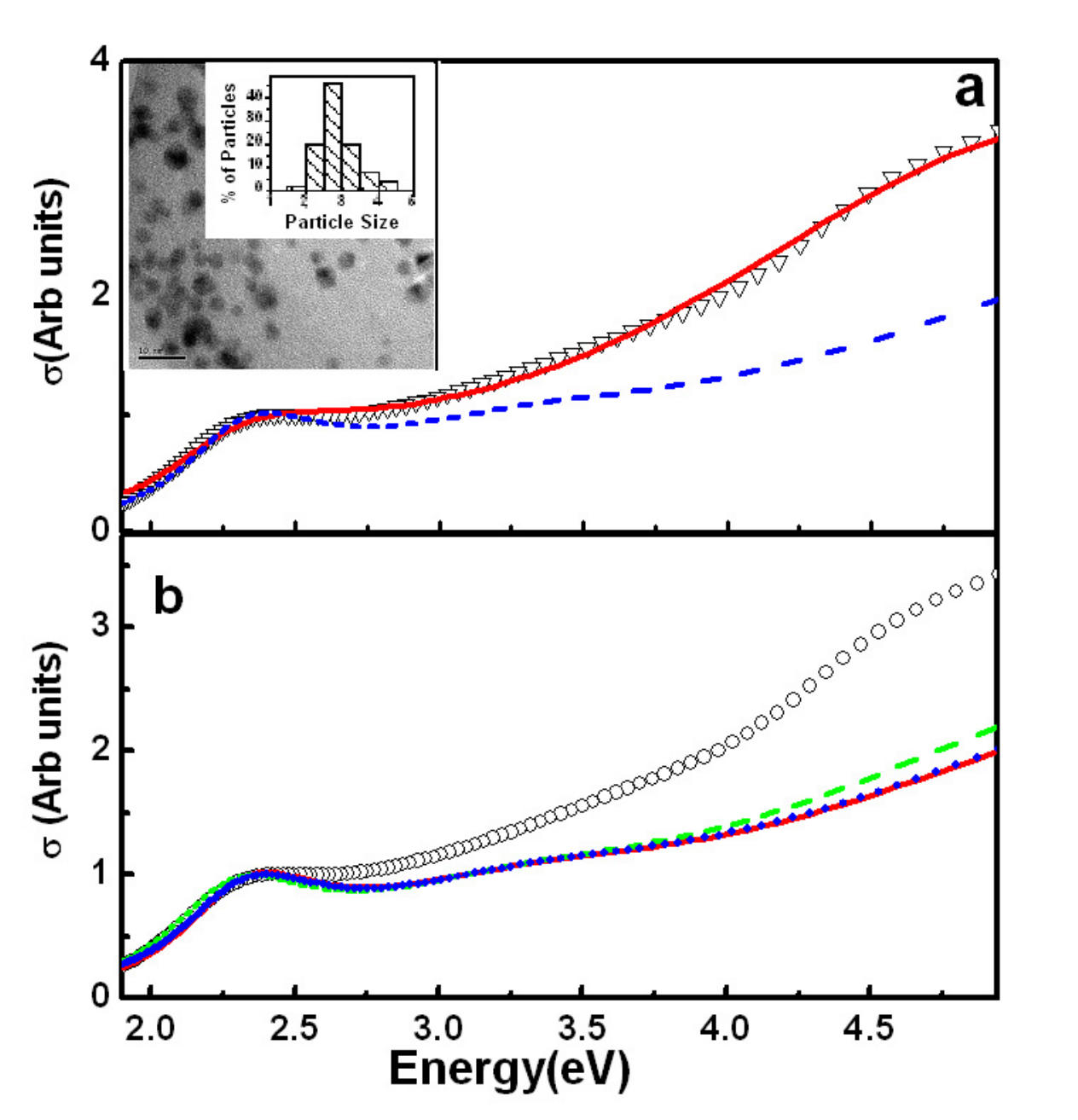}
\caption{ a) Absorption Spectra for particles with diameter 3 nm ($\bigtriangledown$), fitted curves by varying parameters only in  the free electron  (Doted line) and total (interband + free electron) dielectric constant (Solid line). It is clear that varying both interband and  free electron part gives much better fit to the data. Inset: TEM image of a typical gold nanoparticle sample indicating reasonably monodispered particles. b) Spectra for particles with diameter 3 nm ($\medcirc$). Simulated spectra for 3 nm size particles varying only $\omega_p$, $\emph{A}$ ( Solid line ), Varying $\omega_p$, $\emph{A}$ and considering polydispersity ( Dotted line ) and considering coreshell model with a fitted coating thickness of 10 nm using same $\omega_p$, $\emph{A}$ as above ( Dashed line ) . }
\label{fig:1}       
\end{figure}
\begin{table}
\caption{Fitted parameters from Eq. 7 for various values of $\mu$ and r. While the obtained $\omega_1$ and $\gamma_1$ values are almost same as bulk gold, $\omega_2$ and $\gamma_2$  varies significantly with $\mu$ or r.}
\label{tab:1}       
\begin{tabular}{llllll}
\hline\noalign{\smallskip}
$\mu$ &r &$\omega_1$ &$\gamma_1$ &$\omega_2$ &$\gamma_2$ \\
(arb units) &nm & ( eV ) & ( eV ) &( eV ) & eV \\
\noalign{\smallskip}\hline\noalign{\smallskip}
bulk & - & 2.65 & 0.61 & 3.75 & 1.34 \\
1 & 2.47 & 2.61 & 0.57 & 3.84 & 1.38 \\
2 & 2.90 & 2.58 & 0.65 & 3.70 & 1.84 \\
4 & 2.35 & 2.60 & 0.55 & 3.42 & 1.62 \\
7 & 1.40 & 2.67 & 0.72 & 3.26 & 1.86 \\
10 & 1.70 & 2.53 & 0.70 & 3.09 & 1.47 \\
13 & 1.40 & 2.57 & 0.82 & 2.98 & 1.61 \\
16 & 1.50 & 2.56 & 0.77 & 2.90 & 1.19 \\
\noalign{\smallskip}\hline
\end{tabular}
\vspace*{0.5cm}  
\end{table}
\begin{figure}
\includegraphics[scale=0.5]{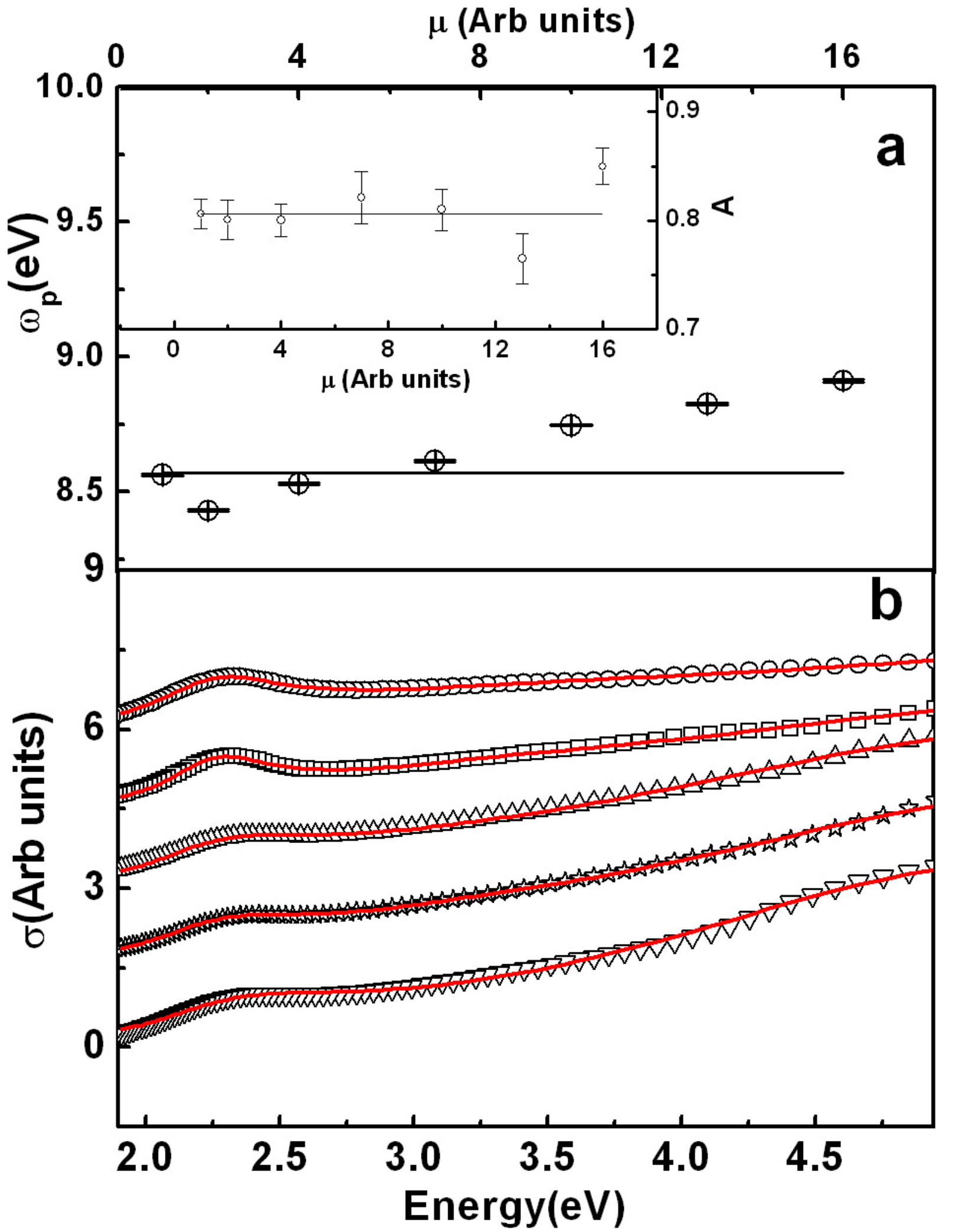}
\caption{ a) Variation of $\omega_p$ values with respect to PMMA capping concentration. Inset : Variation of fitted $\emph{A}$ values with respect to PMMA capping concentration b) Absorption spectra of PMMA capped nanoparticles with various capping concentrations of  $\mu$=1 ($\medcirc$), 4 ($\Box$), 10 ($\bigtriangleup$) 13 ($\star$) and 16 ($\bigtriangledown$) in arbitrary units. Solid line represents fits to the respective curves, using Eq. 6 in Eq. 3.}
\label{fig:fig2}
\end{figure}

The interband contribution to dielectric function in noble metals is dominated by the outermost d electrons in the atoms. The observed peaks in the imaginary part of the dielectric function of gold is due to transitions from the 5 d bands to the 6 s p like conduction bands. The peak at 2.89 eV corresponds to transition from the top of the 5 d band (band 5) to the 6 s p band, while the peak at 4.25 eV corresponds to sum of two different contributions , one from lower lying 5 d band (band 4) to the 6 s p like conduction band and the other between  s p like states in band 6 and 7\cite{36,37}. 
\begin{figure}
\includegraphics[scale=0.2]{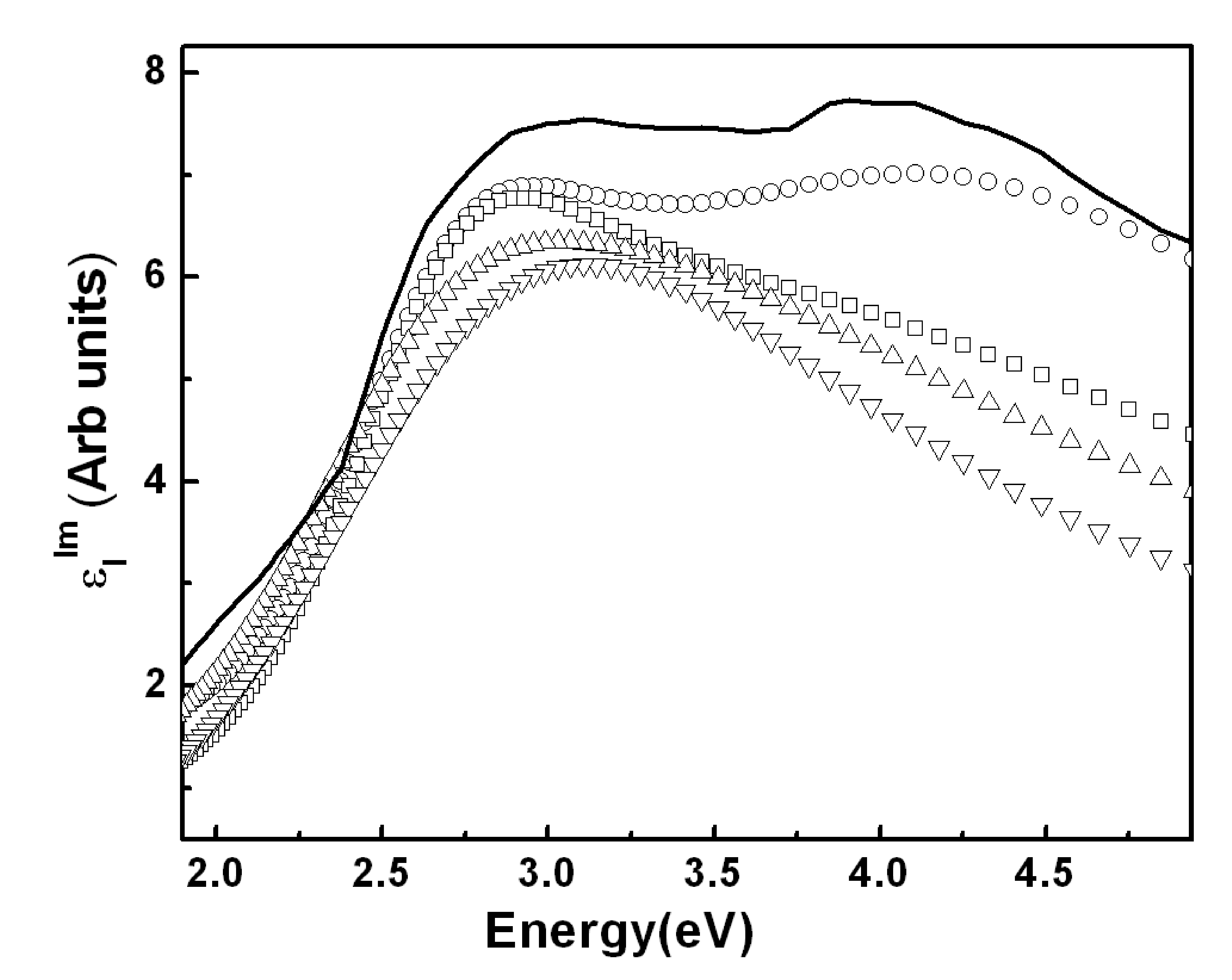}
\caption{  Imaginary part of fitted interband dielectric function. PMMA capping concentrations are   $\mu$=1 ($\medcirc$), 4 ($\Box$), 10 ($\bigtriangleup$) and 16 ($\bigtriangledown$) (refer text for details). Solid line represents Johnson and  Christy value for bulk gold. The curves have been shifted vertically for clarity.}
\label{fig:fig2}
\end{figure}

\begin{figure}
\includegraphics[scale=0.25]{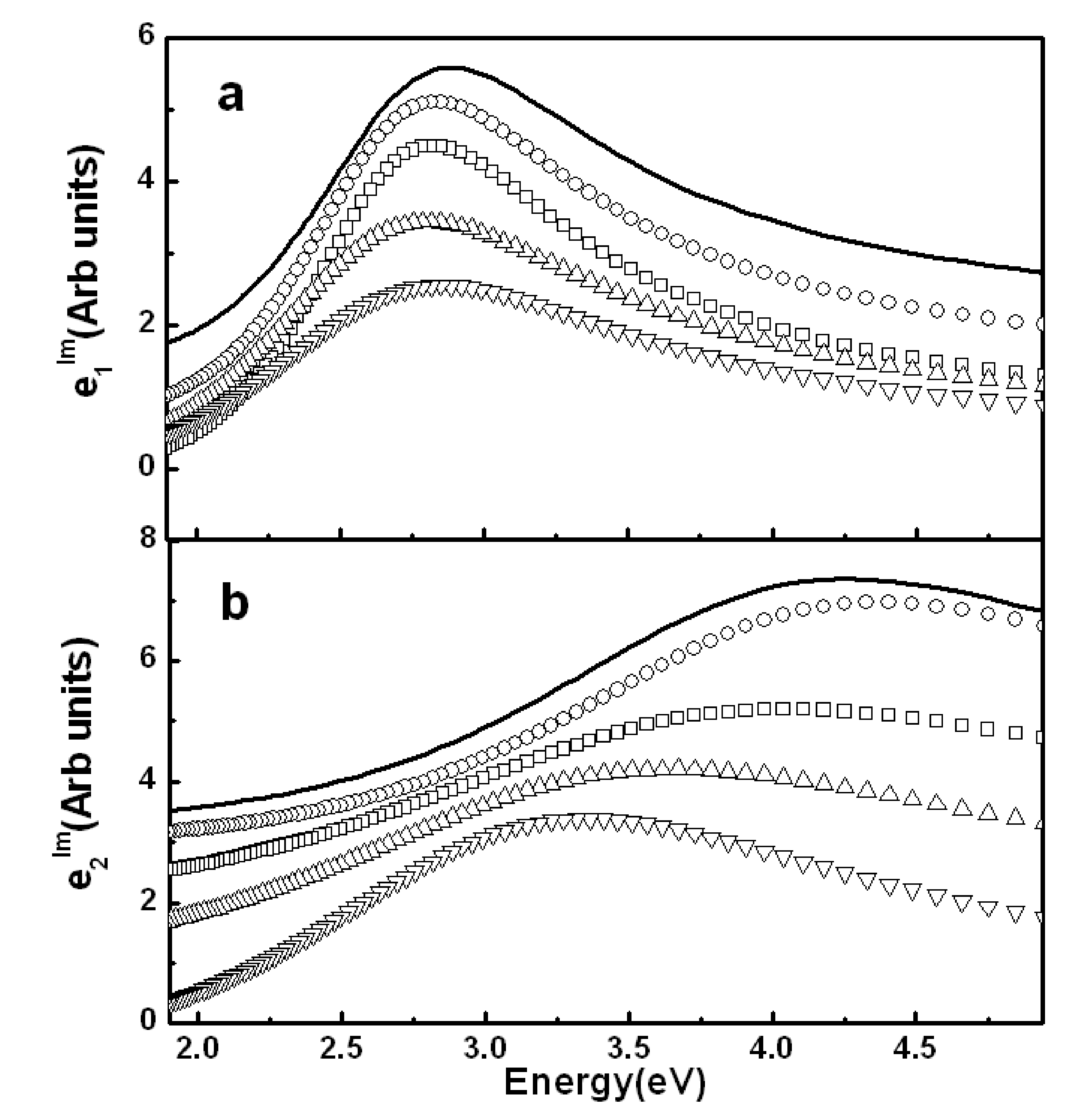}
\caption{ (a) e$_1^{Im}$ and (b) e$_2^{Im}$ . PMMA capping concentrations are  $\mu$=1 ($\medcirc$), 4 ($\Box$), 10 ($\bigtriangleup$) and 16 ($\bigtriangledown$) . Solid lines represents respective calculations using parameters for bulk gold. It is clear that transition corresponding to $\omega_1$ remains same as bulk gold. However, the peak corresponding to  $\omega_2$ shifts towards lower energy  for all particles except those with  $\mu$= 1  for which there is a blue shift compared to bulk gold. }
\label{fig:fig3}
\end{figure}

\begin{figure}
\includegraphics[scale=0.3]{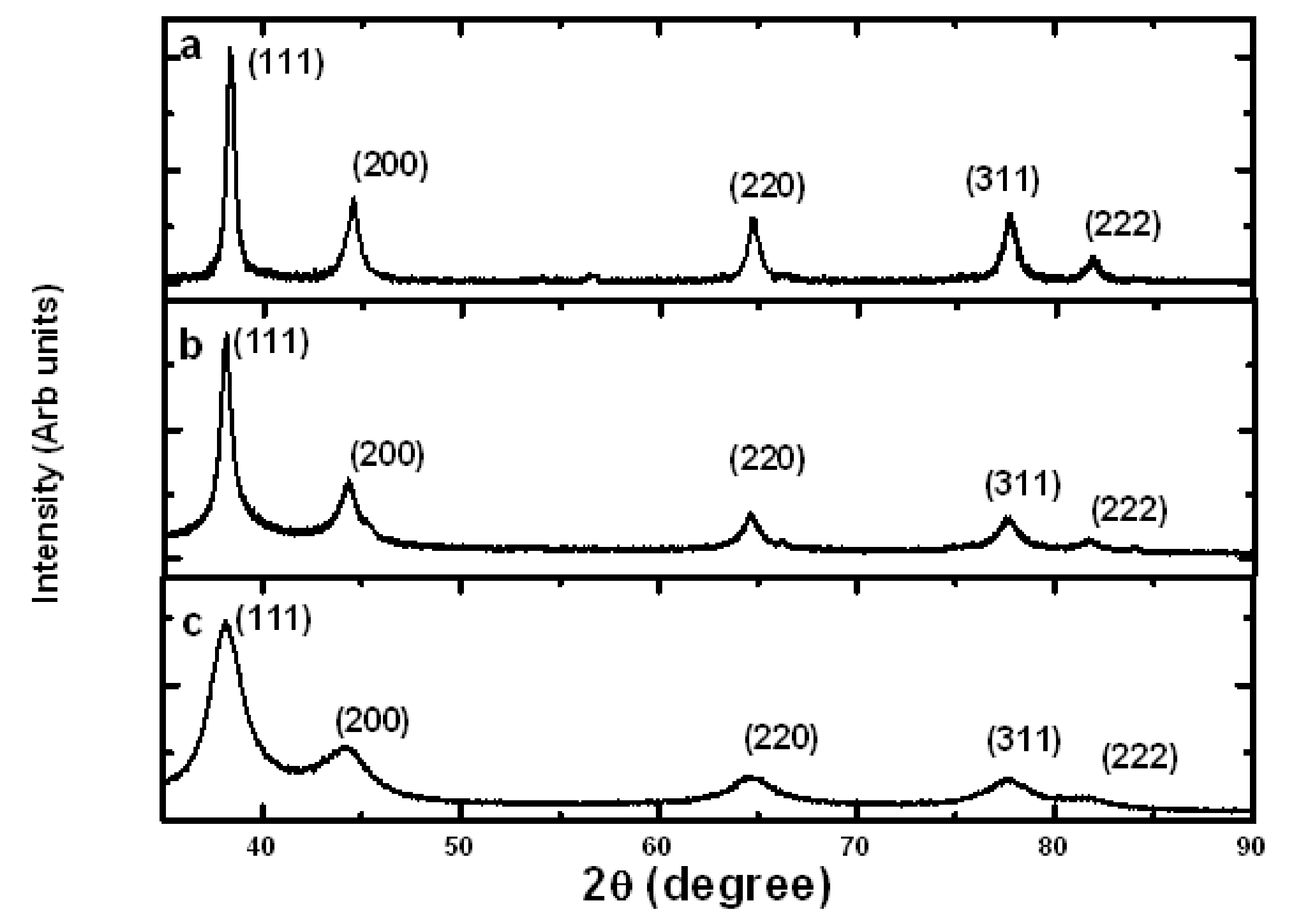}
\caption{XRD Pattern for gold nanoparticles. Data for particles with radius  a) 2.5 nm, b) 2 nm and c) 1.4 nm. The broadening of peaks with decreasing particle size and the shift in peaks for 5 nm particles as compared to the other particles is clearly visible.}
\label{fig:2}
\end{figure}

\begin{figure}
\includegraphics[scale=0.25]{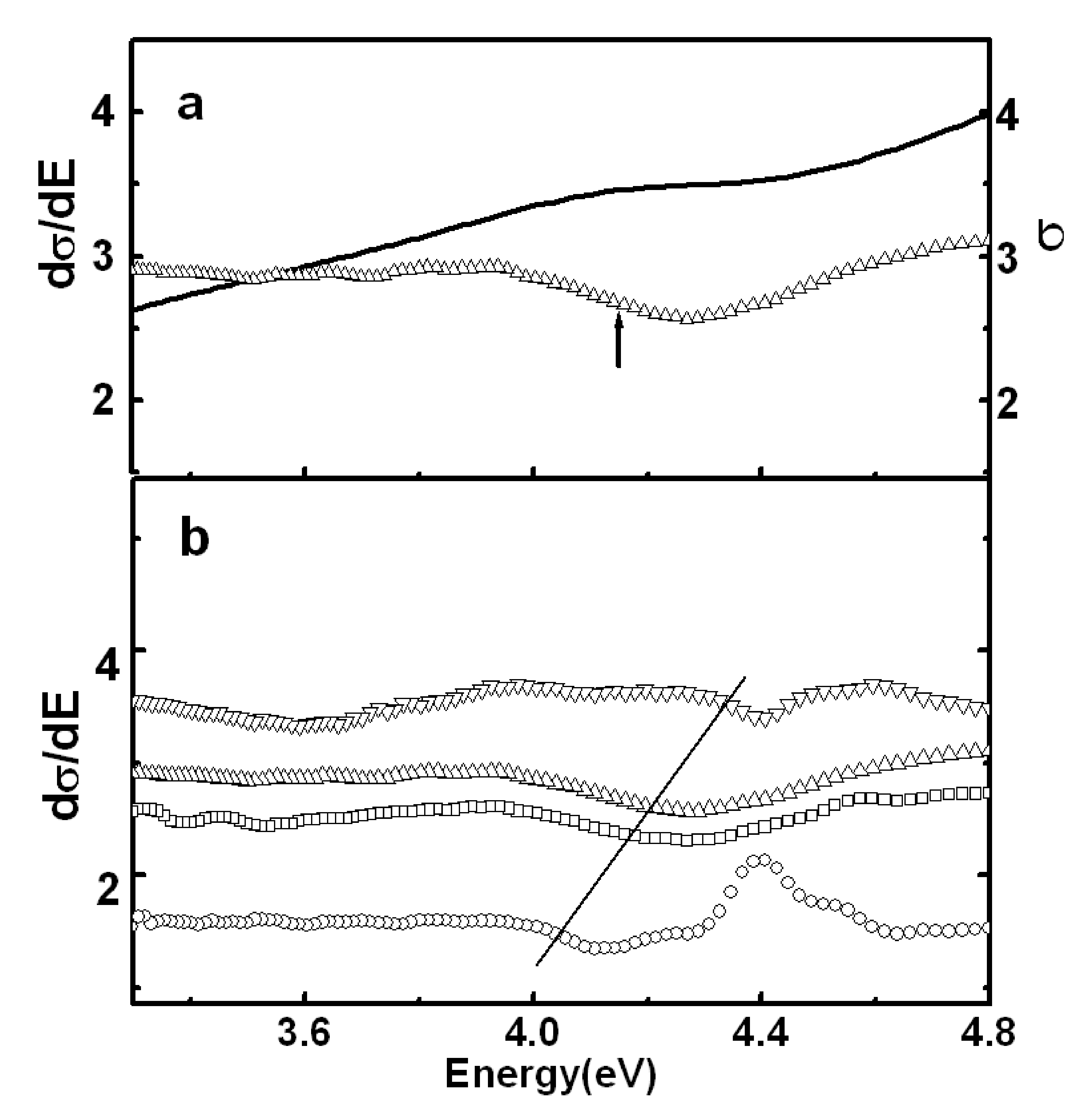}
\caption{ Derivative of absorption data. a) d$\sigma$/dE ($\bigtriangleup$) and $\sigma$ (solid line) for particle with $\mu$= 2.  The arrow indicates position corresponding to peak in e$_2^{Im}$.  b) d$\sigma$/dE for particles with $\mu$= 1 ($\bigtriangledown$), 2   ($\bigtriangleup$), 4 ($\Box$) and 7 ($\medcirc$) in arbitrary units. The line drawn through each of above the data indicates the  red shift in location  of the corresponding  position with increasing concentration,$\mu$. It should be noted that the peaks corresponding to $\omega_2$ = 3.75 eV appears at 4.25 eV, because of the presence of damping constant  $\gamma_2$ in Eq . 3.  The curves has been shifted vertically for clarity.}
\label{fig:epsart}
 \end{figure}

 In general the expression for $\epsilon_I^{Im}$ may be written as
 $\epsilon_I^{Im}$=$\Sigma$$\epsilon_I^{im}$(i-f)  
  where (i-f)represents the transition between initial band i to final band f. The band by band contribution between initial and final state bands are given by $^{21}$ 
\begin{equation}
\epsilon_I^{im}(i-f)=\frac{\rm e^2}{\rm 3\pi m^2\omega^2}\smallint d^3k \vert p_{fi}\vert^2\delta E    
\end{equation}
where E=(E$_f$(k)-E$_i$(k)-$\hbar$$\omega$), p$_{fi}$ is the momentum matrix element corresponding to transition from E$_i$ to E$_f$, e is charge of electron and m the mass. This in principle allows us to identify possible modifications in band structure by studying $\epsilon_I^{Im}$.

\begin{figure}
\includegraphics[scale=0.22]{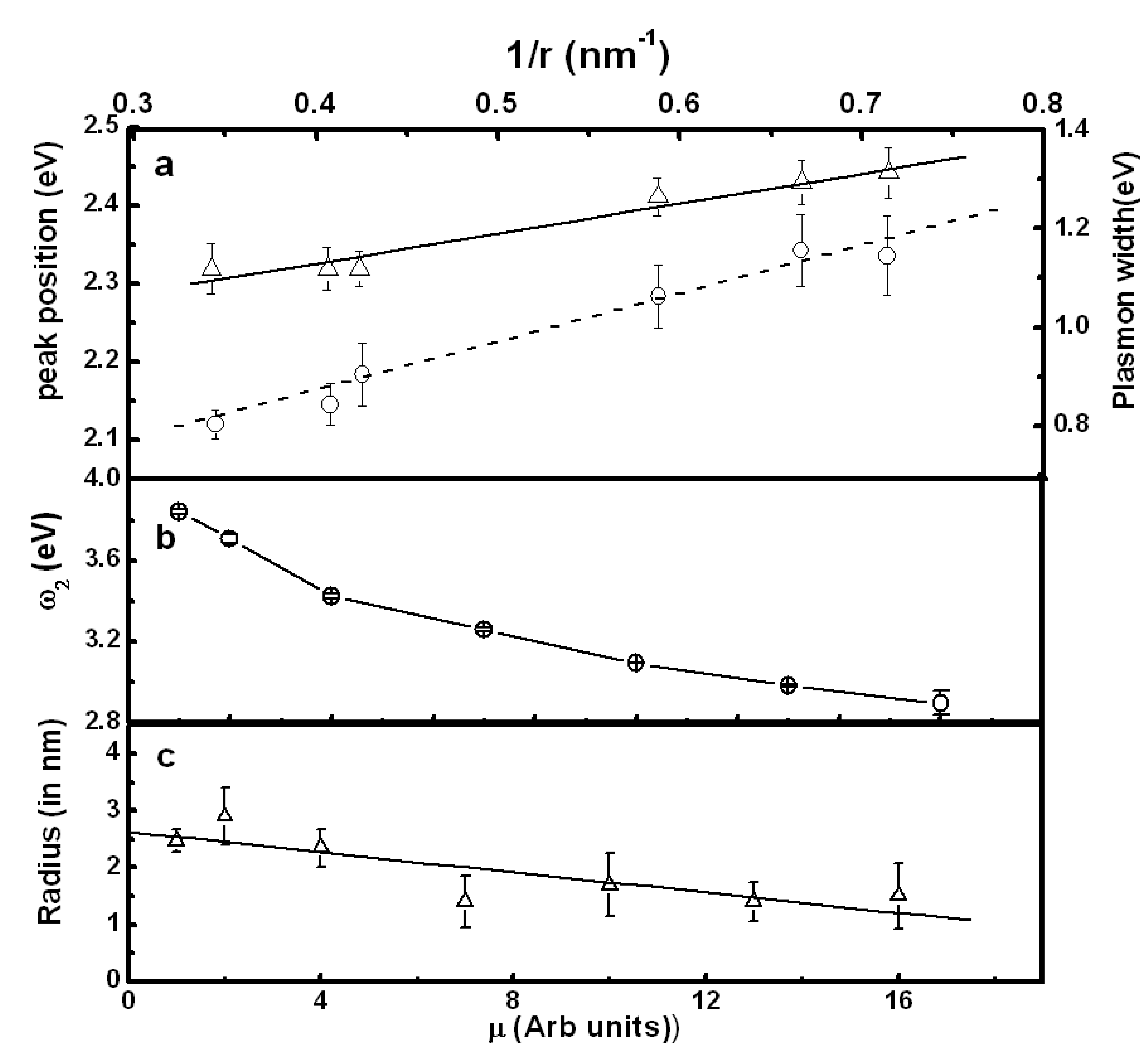}
\caption{ a)  Variation of plasmon resonance peak position ($\bigtriangleup$)and width ($\medcirc$) with increase of gold nanoparticle radius. The blue shift in peak position and increase in width with reduction in particle size is clearly visible. Solid line and dashed lines indicates roughly linear variation of peak position and width, respectively. b) Extracted $\omega_2$ ($\bigtriangleup$)  values for particle with different PMMA concentration. c) Variation of nanoparticle size ($\bigtriangleup$)with PMMA capping concentration. The straight line indicates general trend of size reduction with increasing capping concentration. }
 \label{fig:fig9}
\end{figure}

\section{Results}
\label{result}
  It is clear from Fig. 1(a) that, including only the size dependent damping (Eq. 2) in the free electron part, and keeping the interband part ($\epsilon_I$) of dielectric function fixed at bulk value of gold, as given by Johnson and Christy, does not match the data quite well. We also tried to analyse the data by including various other effects like polydispersity and using a coated sphere model in the Mie calculations, as shown in Fig. 1(b). However it is quite clear that polydisperisty, to the extent it is present in our nanoparticle samples, as well as the combined variation of $\omega_p$, $\emph{A}$ and the coating thickness alone does not modify the calculated absorbance significantly nor does it bring the calculated profile closer  to our data. For simplicity, we have therefore ignored the effects of polydispersity and coating in our model calculations. It turns out that if we use Eqn. 7 and vary both the free electron ($\omega_p$, $\emph{A}$) and interband ($\omega_1$, $\omega_2$, $\gamma_1$, $\gamma_2$) term, then the calculated absorbance matches extremely well with our data in the entire range of measured energy as shown in Fig. 2(b). The obtained values of $\omega_p$ and $\emph{A}$ as a function of $\mu$ is shown in Fig. 2(a). The values of $\emph{A}$ are  almost independent of capping concentration $\mu$ and is found to be 0.8 $\pm$ 0.05 . Interestingly $\omega_p$ seems to increase at higher $\mu$ and this effect will be discussed later. \linebreak

The imaginary part of interband dielectric function, $\epsilon_I^{im}$, obtained from the above analysis  are shown in Fig. 3. The radius of particles, determined from TEM  measurement, was used in all calculations.  It is clear that the peak in $\epsilon_I^{im}$, corresponding to $\omega_1$ = 2.65 eV is almost independent of the size of nanoparticles and is almost equal to the value for bulk gold. However, the peak at $\omega_2$ = 3.75 eV shows strong size and capping polymer concentration dependent effects. This becomes even clearer by considering the obtained values of the fit parameters in Table I. and is highlighted in Fig. 4, where e$_1^{Im}$ and e$_2^{Im}$, as defined in Eq. 8, is plotted. To understand the variation in $\epsilon_I^{Im}$ and especially  e$_2^{Im}$, it is necessary to consider the competing effects of reduced surface atomic co-ordination and lattice contraction on  band structure. XRD measurements, shown in Fig. 5, on powders of nanoparticles with various capping PMMA concentrations have been used to evaluate the changes in the lattice parameter in these particles. These powders were extracted from nanoparticle solutions by controlled evaporation, as described earlier \cite{26}. The FCC symmetry of bulk gold is clearly evident. The calculated value of lattice constants for these three particles are ( $r$ = 1.4 nm , a = 4.082 $\AA$ $\pm$ .005, $r$ = 2 nm , a = 4.085 $\AA$ $\pm$ .006 and $r$ = 2.45 nm, a = 4.062 $\AA$ $\pm $.004 ). There seems to be a lattice contraction of 0.5 $\%$ in the 5 nm diameter particle, while for the smaller particles studied here, the lattice parameters remains unchanged from bulk values. This is a slightly surprising result as compared to the observed size dependent lattice contraction and points to the crucial role of capping polymer in controlling the morphology of the nanoparticles. We also observed size - dependent blue shift of surface plasmon resonance peak position and broadening as found earlier\cite{30,32,33}. The validity and reliability of the scheme of analysis is further exemplified if one looks closely at the derivative spectra of absorbance data, as shown in Fig. 6. One can find that the position in the derivative spectrum indicated with an arrow in Fig. 6(a) matches very closely with the location of peak in the absorbance as well as of the corresponding extracted e$_2^{Im}$. The corresponding positions in the derivative spectrum for a few representative data are indicated in Fig. 6(b). A clear red shift in this position with increasing concentration (decreasing size) is observed (indicated by the solid line), which is similar to the trend in obtained $\omega_2$ values as indicated in Table I.  \linebreak

\section{Discussions}
\label{discus}
In Fig 7(a) we have shown that the plasmon resonance peak for PMMA capped gold nanoparticles blue shifts and broadens with decreasing particle size. This trend has been observed earlier for gold nanoparticles, although in some cases the peak has been found to red-shift as well, depending on embedding matrix and preparation conditions of the nanoparticles. The plasmon resonance broadening has been explained earlier using a model of additional damping due to increased surface scattering of the free electron\cite{7}. This should lead to a 1/r dependance of plasmon resonance width, which we also observe. However the shift in plasmon peak position is much more controversial\cite{10,11,12}. We feel that our results are consistent with a model proposed by Liebsch\cite{10}, where by the strong  screening of the free electron by the polarisable d-band electron in gold in particular and noble metals in general, is effectively reduced due to the quantum mechanical \emph{spill out} effect of conduction band electrons which feels an effectively lower screening and hence oscillates with higher frequency as compared to the electrons confined within the bulk of the nanoparticle. This effect is enhanced in smaller particles, due to increase in surface to volume ratio. Thus observed  trends in the plasmon resonance features provide insight in to the size - dependent changes in the free - electron part of the dielectric function of gold nanoparticles. Fig. 2(a)  also shows variation of plasmon frequency, $\omega_p$ with increase in capping concentration or reduction in particle size. We find that the plasma frequency, generally, increases with increase in capping polymer concentration, similar to the trend of increase in plasmon peak position shown in Fig. 7(a).  However no clear understanding of the changes  occuring in the valence band, especially that in the d band, with reduction in size of nanoparticles, can be obtained unless detailed analysis of the data in the UV region is performed. Using the analytical model for the dielectric function ( Eq. 6 ), we have performed very careful analysis of the entire absorption spectra, including the region from 250 - 500 nm (2.48 - 4.94 eV ) which is dominated by interband absorption in gold. The dielectric constant corresponding to interband transition in gold, $\epsilon_I$ shows two clear peaks, corresponding to transitions from top of 5 d to 6 s p band ( $\omega_1$ = 2.65 eV ) and from bottom of 5 d band to 6 s,p band ( $\omega_2$ = 3.75 eV ), as mentioned earlier. Interestingly our analysis shows that peak corresponding to  $\omega_1$ in $\epsilon_I^{Im}$ remains very similar to that of bulk gold, while that corresponding to $\omega_2$, shifts considerably as compared to bulk values, with increasing capping PMMA concentration or decreasing particle size. This is plotted in Fig. 7(b). We are then led to the conclusion that quantum effects in our gold nano particles leads to strong modification of lower lying  5d bands while leaving the top of the same band relatively undisturbed. Two major factors could be responsible for the observed shifts - reduced surface co- ordination and lattice contraction. It has been discussed earlier\cite{38} for thermal evaporated uncapped gold clusters that lattice contraction leads to broadening of the d bands and increase in their binding energy. On the other hand reduced surface co - ordination leads to opposite effect i.e band narrowing and reduction of d-band binding energy. Remarkably, we find that for particles with lower capping concentration ($\mu$ = 1 ), as well as the largest size, the $\omega_2$ value blue shifts as compared to bulk and there is also a small lattice contraction. Thus it can be concluded that the observed blue shift is largely due to the lattice contraction which overcomes the expected red shift due to reduced co - ordination. However, for all other particles , with smaller size, we find a consistent decrease in $\omega_2$ value, as shown in Fig. 7(b), and no change in lattice parameters. Hence , this would suggest that the reduction in $\omega_2$ values with reduction in particle size is, predominantly due to reduction in surface atomic co - ordination number. Here we would like to point out the cruicial role played by the capping polymer in determining the optical and electronic properties of the gold nanoparticles. Since for the chosen polymer there is no specific interactions with the gold nanoparticles, the nanoparticle polymer interface is quite diffuse. This enables substantial reduction of strain on the small nanoparticles surface, which prevents any significant lattice distortion as is observed for uncapped gold nanoparticles or particles capped with small molecules binding strongly to the gold surface. Table I and Fig. 7(c) shows that the size of the nanoparticles decreases with increasing concentration as was found earlier\cite{26,40}. However if one observes the trend in variation of $\omega_2$ with size and capping concentration, it is clear that even for approximately similar sizes, the difference in $\omega_2$ is quite significant. It is fair to assume that finite size and surface coordination effects must be same for similar sized particles. Hence the  variation in $\omega_2$ for particles of nearly same size but different PMMA capping concentration must be due to variations in surface morphology, which leads to additional changes in surface gold atom coordination. The sensitivity of optical properties of metal nanoparticles to surface morphology has been discussed earlier \cite{39}. Our analysis points to the importance of both quantum and surface effects in determining optical and electronic properties of polymer capped gold nanoparticles. It has also been observed earlier \cite{35,41} that the extend to which the band structure changes in nanoparticles depend on energy level. Normally it has been observed that the shift in position of energy bands closer to the fermi level is less than deeper lying bands\cite{35,41}. This explains why the transition corresponding to $\omega_1$, which arises from top 5 d bands to 6 s p band is almost unchanged from bulk, while transition corresponding to  $\omega_2$ (bottom of 5 d to 6 s p) changes significantly.    \linebreak

Thus it is clear that size - dependent changes in band structure of capped metal nanoparticles is an extremely interesting and complex field of research due to the presence of several competing effects which determines the ultimate properties. Our new scheme of analysis shows how from simple UV- Vis optical absorption spectroscopy one can extract detailed information about changes in both the conduction and valence band in metal nanoparticles due to quantum and surface effects. However, more direct measurements using x - ray spectroscopy might lead to complementary information about size - dependent quantum modifications to the band structure of metal nanoparticles in general, and noble metals in particular.

\section{Conclusion}
\label{concl}    
 Optical properties of polymer (PMMA ) capped gold nanoparticles, in solution has been studied. The systematic variations of the surface plasmon resonance, typical of noble metal nanoparticles and clusters, as a function of nanoparticle size, consistent with some earlier observations is evident. Using a new scheme of analysis, based on modified Mie theory, we have shown how the variations of both the free electron and interband dielectric function of nanoparticle, can be extracted. Careful analysis of the spectra, over a large range of wavelengths, reveals strong sensitivity to the modification of the band structure in nanoparticles, as compared to bulk gold. The optical properties and hence the band structure is very sensitive to various parameters like lattice constant , atomic co - ordination, surface electronic wave function, and nanoparticle - polymer interface morphology. This also highlights the importance of going beyond effective medium theories in nanoparticle - polymer hybrid systems to obtain a deeper understanding of the optical properties of such materials. Further work is in progress to obtain detailed correlation between polymer - nanoparticle surface morphology and its optical and electronic properties. We also plan to perform direct spectroscopic measurements of band structure as well as extending the optical measurements over a larger size range and for various nanoparticle - polymer hybrid systems to understand the generality of the observed behaviour. 

\section{ Acknowledgment}
\label{acknowledge}
 We would like to acknowledge ISRO-IISc, Space Technology Cell for providing financial assistance for this work. (M.H) acknowledges UGC for financial support. We thank DST - IISc Nanoscience Initiative for providing acess to TEM facility.


\end{document}